\def\year{2021}\relax
\documentclass[letterpaper]{article} 
\usepackage{aaai21}  
\usepackage{times}  
\usepackage{helvet} 
\usepackage{courier}  
\usepackage[hyphens]{url}  
\usepackage{graphicx} 
\usepackage{natbib}  
\usepackage{caption} 
\title{Perspective-taking to Reduce Affective Polarization on Social Media}
\author{
    Martin Saveski\footnote{Authors contributed equally.}, 
    Nabeel Gillani\footnotemark[1], 
    Ann Yuan, 
    Prashanth Vijayaraghavan, 
    Deb Roy \\
}

\affiliations{
    Massachusetts Institute of Technology \\
    \{msaveski, ngillani, annyuan, pralav, dkroy\}@mit.edu
}

\usepackage{amsmath}

\usepackage{color} 
\usepackage[dvipsnames]{xcolor}

\begin{document}

\maketitle

\begin{abstract}
The intensification of affective polarization worldwide has raised new questions about how social media platforms might be further fracturing an already-divided public sphere.  As opposed to ideological polarization, affective polarization is defined less by divergent policy preferences and more by strong negative emotions towards opposing political groups, and thus arguably poses a formidable threat to rational democratic discourse.  We explore if prompting perspective-taking on social media platforms can help enhance empathy between opposing groups as a first step towards reducing affective polarization.  Specifically, we deploy a randomized field experiment through a browser extension to 1,611 participants on Twitter, which enables participants to randomly replace their feeds with those belonging to accounts whose political views either agree with or diverge from their own.  We find that simply exposing participants to ``outgroup'' feeds enhances engagement, but not an understanding of why others hold their political views.  On the other hand, framing the experience in familiar, empathic terms by prompting participants to recall a disagreement with a friend does not affect engagement, but does increase their ability to understand opposing views.  Our findings illustrate how social media platforms might take simple steps that align with business objectives to reduce affective polarization.
\end{abstract}

%
%
\section{Introduction}
Political polarization in America is growing more severe~\cite{pewPolarization}, with a ``primacy of partyism'' and strong negative emotions increasingly characterizing inter-group discourse~\cite{primacyPartyism}.  This ``affective polarization''~\cite{affectNotIdeology} is particularly concerning because partisan affect often only weakly correlates with actual policy preferences~\cite{iyengarAffectOrigins}.  For these reasons, affective polarization poses a danger to the democratic process because partisan voters are more likely to sidestep rational discussion and debate when challenging opposing views and crediting their own.

Some scholars highlight homophilic self-sorting and geographic balkanization~\cite{sunsteinRepublic, bishopBigSort} as accelerants of affective polarization because they limit exposure to alternative viewpoints.  The meteoric rise of social media platforms and internet-based political discourse has further fueled self-sorting through the creation and ossification of ``echo chambers''~\cite{sunsteinPolarization,bakshyExposure}, despite the potential these digital tools have to democratize access to diverse social and political perspectives~\cite{shirkyPoliticalPower}.

While much research has explored how echo chambers on social media platforms fuel polarization~\cite{bakshyExposure,polarizationTwitter} or provided theoretical explorations around how connections may be formed between individuals with opposing views in order to reduce controversy~\cite{garimellaControversy}, there have been few experimentally-driven efforts to empirically understand how we might mitigate it.  The few randomized experiments that have been carried out have largely explored whether or not exposure to diverse content---as opposed to exposure to people with diverse political views---reduces political animosity, even though prior research has highlighted the importance of contact between people in opposing groups to reduce inter-group conflict~\cite{contactTheory}.  For example, several digital tools emerged following the 2016 U.S. Presidential Election to enable individuals to explore their own social media echo chambers~\cite{politecho}, read news from politically diverse sources~\cite{readAcrossAisle}, and even connect face-to-face with members of opposing political groups~\cite{hiFromOtherSide}.  However, to our knowledge, none of these tools evaluated the causal effects of their interventions on reducing political animosity between disparate groups.  

There are a few notable exceptions.  One is a recent randomized field experiment by Bail et al.~\shortcite{bailFieldExperiment}, which revealed that exposing partisan Twitter users to a bot that retweets content produced by political elites from the opposing party can actually exacerbate ideological polarization.  The study's findings echo prior work demonstrating how exposure to politically-discordant content can strengthen one's partisan stance and further entrench political divides~\cite{nyhanPersistentMisperceptions}.  Another is a recent randomized field experiment conducted by Levy~\shortcite{levy2020polarization}, which shows that increasing exposure to counter-attitudinal news on Facebook by prompting users to ``like'' the pages of media outlets with opposing political slants can decrease affective polarization.  The different outcomes of these studies suggest that different definitions of polarization, coupled with subtle differences in how people are exposed to opposing views, can have profound effects on our interpretation of the merits and pitfalls of exposure to diverse perspectives.

In this paper, we build on these prior efforts by designing and deploying a randomized field experiment on Twitter.  Much like Levy~\shortcite{levy2020polarization}, we focus on affective (not ideological) polarization given both its rampant acceleration and the threat it poses to rational democratic processes.  We differ from prior experiments, however, by designing a person-centric experience: one that seeks to scaffold exposure to a more holistic view of people from opposing political camps, instead of only exposing them to opposing content.  This focus is motivated by work in social psychology that highlights the value of ``perspective-taking''---or stepping into someone else's shoes to understand their experiences and worldviews---in promoting cognitive empathy~\cite{selman1980understanding} and reducing intergroup conflict~\cite{saxeEmpathy,vescioPerspective,lammEmpathy,batsonPerspective}.  Our focus on empathy is informed by Feinberg et al.~\shortcite{feinbergEmpathyArguments}, who argue that greater empathy for members of disparate groups is one way to reduce strong negative emotions and promote more productive interactions across political divides.  It is further loosely related to recent work from political science illustrating how more personal appeals, e.g., priming a shared American identity~\cite{levenduskyAmericanIdentity} or correcting misconceptions of opposing groups~\cite{ahlerPartiesHeads} can actually reduce affective polarization and perceived social distance.  Importantly, we focus on promoting cognitive empathy \textit{between} opposing groups, given that increasing empathy \textit{within} groups may actually further fuel partisan conflict~\cite{simas2020empathy}.  

Concretely, we deploy a randomized field experiment through a browser extension to 1,611 participants on Twitter, which enables them to ``step into the shoes of another Twitter user'' by randomly replacing their feeds with those belonging to accounts whose political views either agree with or diverge from their own. We find that exposing participants to opposing feeds increases engagement on Twitter, but not cognitive empathy, measured as a self-expressed understanding of why some people might identify with the views represented in the replacement feed.\footnote{The code needed to replicate our analyses are available at: \url{https://github.com/msaveski/twitter_perspective_taking}}  However, first priming participants in familiar, empathic terms by asking them to recall a time they disagreed with a friend increases self-expressed cognitive empathy when browsing opposing feeds.  We believe our experimental design and findings answer a recent call for more behavioral science-inspired interventions to promote demographic discourse online~\cite{spreen2020discourse}, and in doing so, make the following contributions to researchers and practitioners interested in mitigating affective polarization in digital settings:

\begin{enumerate}
    \item A field pilot that enables participants to experience the Twitter feeds of those whose political views may differ from their own,
    \item An analysis of how exposure to opposing feeds affects platform engagement patterns and perceptions of those on different sides of the political aisle, and
    \item Insights into the influence empathic framing can have on how social media users reason about discordant viewpoints and those who hold them
\end{enumerate}

Below, we review related work, describe the details of our experimental design and analyses, explore key results, and discuss their implications for researchers and practitioners interested in promoting understanding and reducing strong negative emotions across partisan divides online.

%
%
\section{Related work}

There is a vast and growing literature that explores political polarization online.  Many of these studies fall into two overarching groups: empirical and theoretical analyses of political polarization, and randomized experiments that seek to further identify, and perhaps even address, the underlying causal mechanisms of polarized attitudes and discourse.  While an exhaustive review of these topics is out of scope for this paper, below we discuss several related studies that ground our work.

\subsection{Analyses of political polarization}
As discussed in the introduction, political polarization comes in many flavors: while ideological polarization describes the extent to which views about specific policies and issues align with one's political group~\cite{pewPolarization}, affective polarization describes the extent to which opposing groups feel strong negative emotions for one another~\cite{iyengarAffectOrigins}. Much of the empirical work exploring political polarization online has explored a concept that is related to, but distinct from, these concepts: political homophily~\cite{birdsFeather}. Political homophily describes the tendency for partisan actors to connect with and consume content from people and sources that reflect their political views. It has likely been of interest to social media and web researchers in part because it is often easier to measure (e.g., by analyzing social ties and content sharing patterns) than other notions of polarization, which may require a more detailed understanding of how members of certain groups perceive others. While homophily is distinct from polarization, studies have shown that a greater propensity to engage with politically concordant groups and ideas can increase both ideological~\cite{sunsteinPolarization} and affective~\cite{levy2020polarization} polarization. This is one of the reasons we design our study to explore the effects of tools that help mitigate such homophily online---and view doing so as a possible step towards mitigating polarization.

There have been several analyses of political homophily on digital platforms, especially on social media platforms like Facebook and Twitter, where politically-active users generally tend to connect with and consume content from sources that align with their political views~\cite{bakshyExposure,polarizationTwitter,cossard2020vaccines,garimella2021political,cinelli2021echo}. There is also evidence suggesting these trends, at least on Twitter, are intensifying over time~\cite{twitterLongTerm}. As these trends unfold on digital platforms, however, there continues to be a debate about their root causes. On the one hand, several have described the existence of a ``filter bubble'' effect, where personalization algorithms on search, social media, and other platforms increasingly drive people into buckets of other people and content that reflect their worldviews~\cite{pariser2011filter,robertson2018auditing,le2019personalization,dosSantos2020twitter}. The recent experiment by Levy~\shortcite{levy2020polarization} described earlier appears to confirm this hypothesis and illustrates how such selective exposure can increase affective polarization. On the other hand, there is a belief that while platforms play a role in promoting selective exposure through personalized recommendations, these recommendations are ultimately grounded in user preferences and psychology, as evidenced by who they follow and what they consume and share~\cite{bakshyExposure}. Invariably, much more empirical work is needed to better-understand how both homophilous and heterogeneous political interactions manifest online (e.g., as in An et al.~\shortcite{an2019discussions}), what causes them, and how these relate to different types of of polarization.

While homophily may be linked to polarization, there is still debate on how to promote more cross-cutting dialog and exposure to opposing views, if at all. For example, Gillani et al.~\shortcite{socialMirror} and Garimella et al.~\shortcite{garimellaControversy} highlight the importance of thoughtfully selecting accounts that are more likely to bridge opposing viewpoints. Recent studies have also shown that simply using political stance labels in content~\cite{gao2018stance} or priming participants with signals that reveal opposing political views~\cite{centolaSocialLearning} can exacerbate existing divides, highlighting the challenges involved in fostering more exposure and exchanges across partisan groups. Other studies have investigated the role that media outlets play in political polarization online and how they can change the language of their promotional social media posts to engage more politically diverse audiences~\cite{saveski2022engaging}. Researchers have also qualitatively explored attitudes towards facilitating more exposure to diverse perspectives online---seeking to uncover under which conditions, and for which groups, exposure to diverse perspectives may be favorable and productive for achieving democratic aims~\cite{nelimarkka2019redesign,gron2020design}.

\subsection{Randomized experiments to understand polarization}
Our work also builds on recent studies that have used randomized experiments to study the causal effects of exposure to opposing views on ideological and affective polarization. 

Bail et al.~\shortcite{bailFieldExperiment} conduct a randomized experiment incentivizing participants to follow a Twitter bot that retweets posts by influential accounts with opposing political views. While the intervention did not affect Democrats, they found that Republicans expressed even more conservative views after following a liberal bot, suggesting that there might be a ``backfire'' effect. In contrast, Guess and Coppock~\shortcite{guess2020does} run a series of survey experiments exposing participants to opposing views but find no evidence of ``backfire'' effects. A recent study by Spitz et al.~\shortcite{spitz2021interventions} that focuses on the identity of the source of opposing views provides some insights on when ``backfire'' effects are more likely to occur. Using a randomized survey experiment, they find that opposing views expressed by celebrities, liked or disliked, did not affect participants' willingness to change their opinions but that opposing views expressed by experts lead to even further issue polarization. 

Other studies have sought to understand the effect of exposure to partisan news on affective polarization. Levy~\shortcite{levy2020polarization} ran a randomized experiment encouraging participants to subscribe to the Facebook pages of conservative and liberal media outlets. He finds that participants are willing to engage with opposing news on social media and that such exposure leads to a decrease in affective polarization. In a similar experimental design, Guess et al.~\shortcite{guess2021consequences} run a randomized experiment incentivizing participants to set their browser homepage to a left- (HuffPost) or right-leaning (Fox News) news source and subscribe to corresponding Facebook pages. In contrast to Levy, they do not find any significant effects on the participants' feelings toward the parties or their perceptions of polarization. Instead, they find a significant decrease in the participants' trust in mainstream media up to one year later. 

In contrast to previous studies, we design a person-centric intervention that offers participants more holistic exposure to people with opposing political views, instead of only exposing them to opposing content.

%
%
\section{Experimental design and recruitment}
\paragraph{Browser extension.}

\begin{figure*}[t!]
\centering
\includegraphics[width=\linewidth]{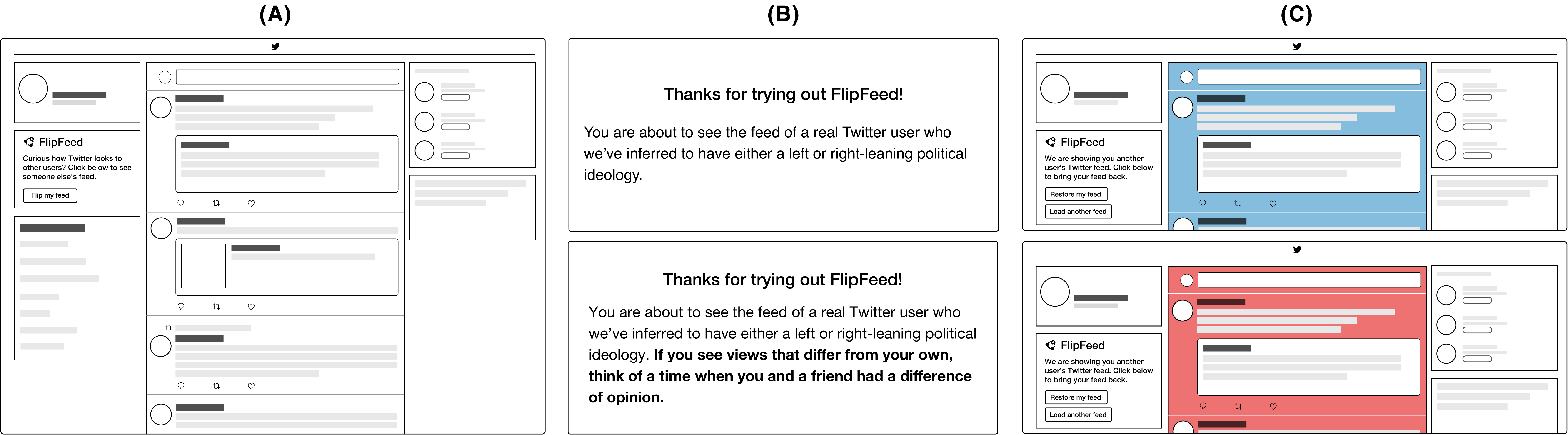}
\caption{\textbf{(A)} Illustration of Twitter's web interface with the Chrome extension installed. Users can press the ``Flip my feed'' button and replace their feed with someone else's Twitter feed. They can flip their feed multiple times or return to their own feed at any point. \textbf{(B)} Control and empathic prompt (added emphasis) shown at random to the users when they first flip their feed. \textbf{(C)} Blue (left-leaning) or Red (right-leaning)  destination feed shown to the users after they press the ``Flip my feed'' button (colors shown only in the illustration).}
\label{fig:ui}
\end{figure*}

Our experimental intervention is deployed through a Google Chrome extension that allows Twitter users to replace the content in their feeds with content from another real left- or right-leaning Twitter account's feed.  After installing the extension, the user presses the ``Flip my feed'' button under their profile picture, which randomly selects one of 200 pre-curated ``destination feeds'' to replace their own.\footnote{To select the accounts, we started with a set of accounts randomly sampled from those that were active in political discourse on Twitter during the 2016 U.S. Presidential Election cycle~\cite{lsmElectome}.  Three human annotators independently classified each account's political ideology as either left or right-leaning.  We selected the first 100 left and right-leaning accounts that each of the annotators agreed on, respectively, to arrive at our set of 200.} These destination feeds are approximate chronological reconstructions of a real Twitter user's feed in order to cultivate an experience of seeing Twitter from another user's point of view with the highest possible fidelity.  The user can then browse and interact (via likes, replies, and retweets) with the new feed as if it were their own.  At any time, the user can load another randomly-selected feed.

Our extension enables users to ``step into someone else's Twitter feed'' to experience their social media universe, and implicitly, reason about the content and other media forces that likely shape their political opinions.  Our overarching hypothesis is that this process can activate the user's theory of mind~\cite{theoryOfMind}, a critical element in the practice of perspective-taking: a practice that has proven to promote empathy and reduce intergroup tension and animosity in other contexts~\cite{saxeEmpathy}.   

\paragraph{Outcomes.}
We measure a series of indicators to better-understand how different aspects of the extension influence users' beliefs and behaviors:

\begin{itemize}

\item \textbf{Usage}: the total amount of time (in seconds) spent browsing destination feeds.

\item \textbf{Engagement}: the number of times participants clicked, liked, or retweeted one of the tweets shown in the destination feed.

\item \textbf{Survey responses}:  after a user scrolls through 50 tweets in a given feed, they are prompted (with 50\% probability\footnote{With the exception of the first time they load another feed, in which case we always show users a questionnaire after they scroll through 50 tweets.}) to respond to the following four questions using a sliding scale from 0 to 100 (0 = Strongly disagree; 100 = Strongly agree):

\begin{itemize}
\item \textbf{Q1}: ``This feed is different from what I'm used to seeing''
\item \textbf{Q2}: ``I learned something new from browsing this person's feed''
\item \textbf{Q3}: ``I can understand why some people might identify with the views shown in this feed''
\item \textbf{Q4}: ``In the future, I would be interested in having a conversation with this feed's owner''
\end{itemize}

\end{itemize}

The above questions serve different purposes:  Q1 and Q2 are meant to validate the effectiveness of the experience in actually exposing people to politically disparate/novel information and people; Q3 seeks to assess the extent to which stepping into someone else's social media universe actually enhances their ability to reason about why someone might hold a particular set of political views; and Q4 seeks to measure to what extent participants might be open to deeper in-person engagement with those who hold opposing political views.

\paragraph{Treatments.}
We use a two-by-two factorial design. The first factor varies the political leaning of the feed served to the user: same or opposing (as inferred using the political alignment metric described in Section~\ref{sec:methodology}).  The second factor varies the type of initial prompt shown to the user: control vs. empathic. Participants are assigned to a prompt condition only once, but are assigned to a different feed condition every time they load another feed. In both cases, participants / sessions were assigned to experimental conditions using Bernoulli randomization with probability~of~0.5. 

Figure \ref{fig:ui} provides a visual depiction of these treatments, which we describe in more detail below.

\begin{enumerate}
\item \textbf{Feed ideology:} Each time a user loads another feed, they are randomly served the feed of one of the 200 pre-selected accounts.  On average, 50\% of the time they are shown a left-leaning (``blue'') feed; the other 50\%, a right-leaning (``red'') feed (Figure \ref{fig:ui}C).  As a result, on average, 50\% of the time participants see a feed that approximately aligns with their political views, and the other 50\% of the time they see one depicting opposing~views.
\item \textbf{Empathic prompt:} Prior work~\cite{saxeEmpathy,transgenderCanvassing,refugee2018Letters} has demonstrated how prompting ``perspective-taking'', i.e., putting one's self in the shoes of an ``outgroup'' member, can reduce feelings of animosity towards that group.  With this in mind, we show the following ``empathic prompt'' to a random 50\% of users right before they load another feed for the first time: ``If you see something you disagree with, think of a time you and a friend had a disagreement'' (Figure \ref{fig:ui}B).  The remaining users are enrolled in the control group, where they only see instructions on how to use the tool.  The motivation for including this treatment is to explore if we can trigger framing effects~\cite{prospectTheory} through the use of familiar, non-adversarial language to help mitigate inter-group animosity.
\end{enumerate}

\paragraph{Participant Recruitment.}
We sent Twitter Direct Messages to 1,600 potential participants in early 2017.\footnote{Only 1.4\% of Direct Message recipients ended up using the extension, suggesting media coverage and word-of-mouth drove most participation.}  Potential participants were sampled from a wider set of 1M Twitter accounts who discussed the 2016 U.S. Presidential Election on the platform between June and mid-September 2016.  Media coverage of the extension also played a key role in driving participation. The nature of our data makes it impossible to know exactly how many participants discovered and used the extension through each recruitment channel. 

We note that all results reported subsequently in the paper are conditional on a person's decision to use the extension, and so, we cannot generalize findings to a wider population (e.g., all Twitter users in the US, or even more broadly, all adult members of the U.S. population). However, we argue that this self-selected group is an important and relevant population to study. A large-scale application of a perspective-taking intervention on social media platforms like Twitter and Facebook is likely to be deployed as a new platform feature that users can choose to use, rather than being imposed upon all users.\footnote{Note that this is different from content-centric interventions (e.g., as in Bail et al.~\shortcite{bailFieldExperiment}) where platforms, with minimal user interface change, can decide to promote or add out-of-network cross-cutting content to the users' feeds.} Furthermore, achieving progress towards our ultimate goals of reducing intergroup animosity and promoting understanding does not require everyone in the population to participate: instead, engaging even a small subset of committed actors may be sufficient.

%
%
\section{Methodology}\label{sec:methodology}
\paragraph{Measuring Political Alignment.}
We infer participants' political alignment using their historical sharing patterns. Previous studies have shown that user's sharing patterns (i.e., which websites they link to in their posts) are predictive of their political alignment~\cite{bakshyExposure}. We collect all tweets posted by each participant one month prior to their first use of the extension and define their political alignment as the average score of the political alignment of the domains linked to in their tweets. The score ranges from -2 (very liberal) to 2 (very conservative). To determine the users' political leaning, we threshold their alignment score at zero. We use the domain political alignment scores published by Bakshy et al.~\shortcite{bakshyExposure} obtained by analyzing the sharing patterns of 10.1M Facebook users who self-reported their political affiliation. Finally, we compared the inferred political leanings based on the users' content sharing patterns with the political leanings inferred from the users’ following patterns using the method proposed by Barber{\'a} et al.~\shortcite{barbera2015tweeting}. We find that the predictions of the two methods agree in 88.62\% of the cases. However, the method based on the users’ content sharing patterns allows us to estimate the political leaning of 40\% more users.

\paragraph{Reconstructing Feeds.}
To obtain the data needed to reconstruct the Twitter feeds of the 200 ``feed giver'' accounts we used in our experiment (100 left-leaning, 100 right-leaning), we run two processes. The first process runs daily and downloads the list of accounts followed by each of the feed givers. The second runs continuously and downloads any new tweets posted by these followed accounts. We used the Twitter REST API to collect this information in a timely fashion.

When a participant loads another their feed, we do the following: ($i$) sample one of the 200 feed givers, ($ii$) fetch the feed giver's followees from the local database, ($iii$) fetch the tweets of the feed giver's followees from the local database, and ($iv$) sort tweets in reverse chronological order. To increase the chances that participants see content from a wide range of accounts, we make sure that no more than three tweets posted by the same author appear sequentially. To ensure that tweets are up-to-date and have not been deleted since they were downloaded initially, we download them again using the Twitter API right before serving them to the participant. Finally, to prevent exposing users to any inappropriate content, we filter out tweets flagged as ``possibly sensitive'' by Twitter. We note that we opted for reverse-chronological ordering of the reconstructed feeds as we were unable to replicate the algorithmic curation used by Twitter to produce personalized feeds.

\paragraph{Model Specifications.}
To estimate the causal effects of the two treatments, we use Bayesian linear mixed-effects models~\cite{gelman2006data, mcelreath2020statistical} which allow us to simultaneously account for two key dependencies in the data: ($i$) that a participant may load another feed multiple times, which will result in multiple correlated observations; and ($ii$) that the same feed giver may be assigned to multiple participants.

To avoid any biases that arise from the flexibility of choosing any model specification, we restricted ourselves to using only the maximal random effects structure that is justified by our design~\cite{barr2013random}.  For analyses with this type of hierarchical structure, it is common to use frequentist linear mixed effects models (e.g., using the \textit{lmer} package in R~\cite{bates2015fitting}). However, we found that the maximal frequentist linear mixed effects models did not converge using this method, likely due to data sparsity, leading us to use their Bayesian counterparts.\footnote{We note that such convergence issues are common when fitting (frequentist) maximal mixed effects~\cite{barr2013random}. The recommendation in these cases is to either use Bayesian approaches or fit simpler models that do converge. As we discuss later, our robustness analyses show that fitting simpler frequentist models leads to very similar results as fitting the full Bayesian models.}

Given the by-participant and by-destination-feed-owner dependencies described above, the model with the maximal random effects structure is:
%
\begin{align*} 
Y_{pg} = \beta_0 & + \beta_1 feed + \beta_2 prompt + \beta_3 (feed \cdot prompt) \\
       & + \beta_{x:feed} (feed \cdot \mathbf{X_p}) + \beta_{x:prompt} (prompt \cdot \mathbf{X_p})  \\
       & + P_{0p} + P_{1p} feed + G_{0g} + G_{1g} prompt + \epsilon_{pg}, 
\end{align*}
where $Y_{pg}$ is the outcome for participant $p$ who was shown user $g$'s feed; $\beta_0$ is the baseline outcome, $\beta_1$ and $\beta_2$ are the main effects of the opposing feed and empathic prompt conditions, respectively; and $\beta_3$ is the interaction effect of the two conditions.  This interaction term allows us to model how the two treatments interact with each other. For instance, one hypothesis is that the empathic prompt might be more effective when the user is shown a feed of someone with an opposite political leaning. 

The coefficients $\beta_{x:feed}$ and $\beta_{x:prompt}$ are associated with the terms that interact the two treatment assignments with the participants' covariates ($\mathbf{X_p}$). We considered five participant covariates: number of posts and number of likes made throughout their history of Twitter use, number of accounts they are following, number of accounts following them, and length of tenure on Twitter.\footnote{We logged and scaled all covariates expect for length of tenure as their distributions were highly-skewed.} We included these covariates to adjust for any imbalance between the treatment and control groups and potentially gain statistical power. The covariates serve as proxies for various indicators of how active and well-known the accounts were on Twitter, which presumably correlate with other off-platform characteristics likely to influence response to treatment like profession, demographics, and other characteristics, which we do not have access to.

The by-participant random effects include random intercepts ($P_{0p}$) and random (feed opposite) slopes ($P_{1p} feed$) that allow the participants' responses to vary according to which feed ideology condition they were assigned to. Similarly, the by-feed-giver random effect include random intercepts ($G_{0g}$) and random (prompt) slopes ($G_{1g} prompt$) that allow the participants' responses to vary according to the prompt condition they are assigned to. 

To fit the models, we use the R package \textit{brms}~\cite{burkner2017brms, burkner2018advanced}, which uses Markov Chain Monte Carlo to perform posterior Bayesian inference on the coefficients and intercepts highlighted above. We specify $\mathcal{N}(0,30)$ priors for the coefficients and $\mathcal{N}(0,1)$ for the standard deviations. We use four chains to estimate each model and run each chain for 10,000 iterations.

\paragraph{Assessing Model Robustness.}
To test the robustness of our results to variations of the model specification we fit eight different model specifications, varying three key aspects of the models: ($i$) whether the model includes a term interacting the two treatment assignments (feed and prompt), ($ii$) whether the model includes the user covariates, ($iii$) whether the model is frequentist (\textit{lmer}~\cite{bates2015fitting})---which is less-than-maximal and includes only random intercept terms for feed giver and participant (instead of random intercepts \textit{and} slopes)---or Bayesian (\textit{brms}~\cite{burkner2017brms, burkner2018advanced}).

In most cases, all eight model specifications lead to the same substantive conclusions. The estimates and credible/confidence intervals (CIs) of the corresponding \textit{brms} and \textit{lmer} models are very close. Similarly, including the covariates in the models leads to very small changes in the estimates. The most consequential modeling decision is whether to include a term that interacts the two treatment assignments. While including an interaction term does not drastically affect the point estimates for the inferred slopes and intercepts, it does lead to significantly wider CIs.  This is particularly the case for the CIs of the prompt term for the survey question responses. We note that the model specification used for our main results (\textit{brms}, with interactions, with covariates) leads to the most conservative results. We omit the results from our robustness checks in the body of this paper due to space constraints, but refer interested readers to Section~\ref{sec:replication-and-supmat} for more information on how to access the supplementary and replication materials.

\paragraph{Ethics and Protection of Human Subjects.}
Our research protocol was approved by the MIT's Institutional Review Board. All participants downloaded and installed the extension through the Google Chrome Extension store.  The store included a link to the project's webpage, which contained information about the extension's purpose, function, and developers.  Furthermore, the webpage contained a link to a privacy policy describing which data would be collected from participants.  Participants could email an alias designated for the project to ask any additional questions, and also, request that their data be removed from the research study at any point.  No such requests were received.  We did not ask for informed consent from participants, nor did we detail the specific research design or motivating research questions, as doing so could introduce bias and compromise the integrity of the study.

%
%
\section{Results}
\begin{figure*}[t]
\centering
\includegraphics[width=0.99\linewidth]{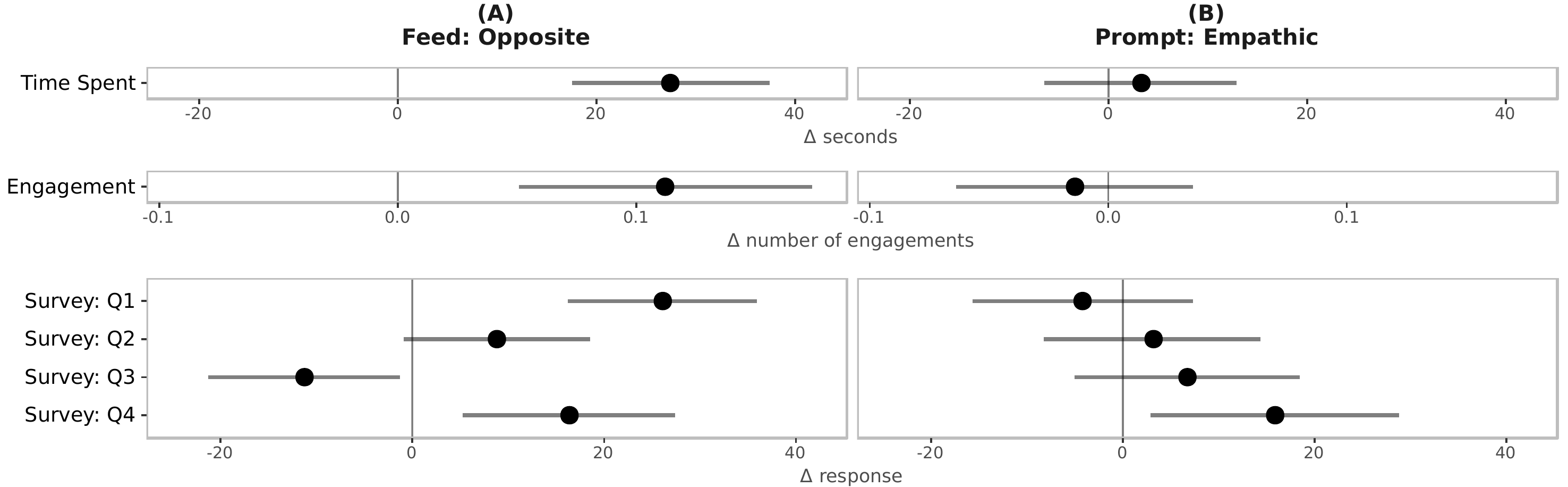}
\caption{Effects of different experimental conditions on users' behavior (time spent and engagement) and answers to survey questions. \textbf{(A)} Showing a feed that aligns with the users' political views (same) vs. one that does not (opposite); \textbf{(B)} Showing the control vs. empathic prompt to the user after they load another feed for the first time.  Error bars show 95\% credible intervals.}
\label{fig:main-effects}
\end{figure*}

\begin{figure*}[t]
\centering
\includegraphics[width=\linewidth]{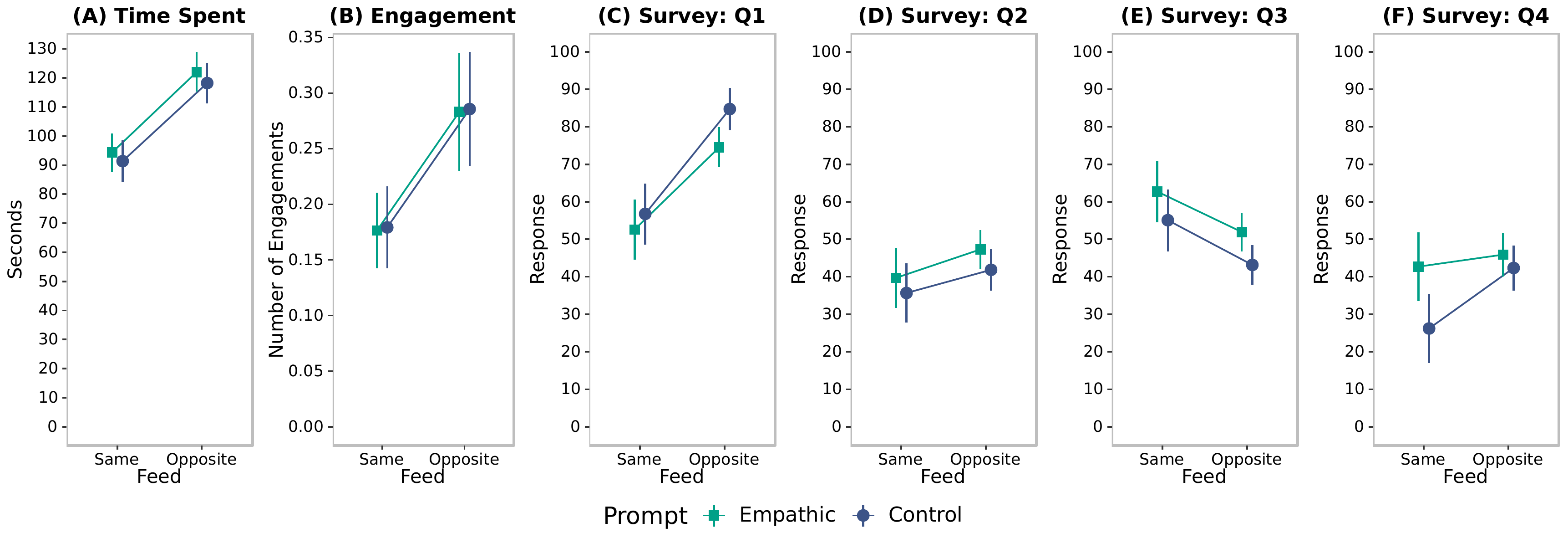}
\caption{Interactions between the effects of the two treatments.  The lack of significant interaction effects in this case means that the slopes of each line within each subfigure are not statistically significantly different from the other.  However, using non-linear hypothesis tests, we find significant differences across some \textit{pairs} of treatments.  In particular, we see that those who see the empathic prompt and an opposing feed are less likely to view the feed as different from what they are used to seeing (C) and more likely to understand why some people might identify with the views represented in the feed (E) compared to when they see the control prompt.  Error bars show 95\% credible intervals.}
\label{fig:interaction-effects}
\end{figure*}

Below, we present the effects of our two main treatments (Same vs. Opposite Feed, Empathic vs. Control Prompt) and their interactions on the engagement and survey measures described earlier. 
Figure~\ref{fig:main-effects} shows the treatment effects of each of the two treatments, and Figure~\ref{fig:interaction-effects} shows the absolute values of the outcomes across the four treatment conditions. 

In total, 1,611 users whose political leaning we could infer participated in the randomized experiment. On average, participants replaced their feed with a feed of another Twitter user 3.14 times. All participants replaced their feed at least once and were not required to do so more than once. They skewed left (91\%), had an average of 5.9 years tenure on Twitter at the time, and a median of 2,039 posts, 736 favorites, 439 followees, and 294 followers. In the online supplementary material, we show that participants assigned to different treatment conditions did not have significantly different characteristics.

%
%

\paragraph{Same vs. Opposite Feed.} Figures~\ref{fig:main-effects} A and B show changes in platform engagement (including time spent browsing feeds) and responses to the survey questions, respectively, as a result of each treatment. Participants spend 29\% more time (27.4 average increase in seconds, 95\% CIs [17.6, 37.4]) browsing and are 37\% more likely to click on or engage with (like, retweet, or reply to) tweets shown in politically-opposed destination feeds (0.11 average increase in engagements, 95\% CIs [0.05, 0.17]), suggesting these feeds enhance engagement. One possibility is that participants are more curious to explore novel information when they browse an opposing feed.

When asked whether a destination feed contains content that is very different from what they are used to seeing on Twitter (Survey: Q1), users who see a politically-aligned feed provide a neutral answer (on average, 58 out of 100), while users who see politically-opposed feeds are more likely to agree (26.1 point increase, 95\% CIs [16.2, 35.9]) that the feed looks very different than what they are used to seeing. These results confirm that our treatment is valid and has the desired primary effect of exposing users to political views outside of their social media bubbles.

Participants who view politically-opposed feeds are also more likely to report that they learned something new by browsing the feed, though this effect is marginally statistically insignificant (Survey: Q2, 8.8 points increase, 95\% CIs [-0.9, 18.5]); however, they are less likely to say that they understand why others may hold such views (Survey: Q3, -11.2 points change, 95\% CIs [-21.3, -1.3]). The exposure to opposing views, however, triggers an interest in learning more about the other side: compared to participants who saw a feed that aligns with their views, participants who see politically-opposed feeds are more likely to report that they would like to have a conversation with the feed owner in the future (Survey: Q4, 16.4 points increase, 95\%~CIs~[5.3,~27.4]).

The results suggest that using social media platforms to simply expose users to those who hold different political perspectives can help increase engagement, but on their own, are insufficient in bridging understanding gaps and promoting empathy between disparate groups.

\paragraph{Empathic Prompt.} Showing participants an empathic prompt does not have a significant effect on usage or engagement: participants spend slightly more time browsing feeds (3.3 seconds change, 95\% CIs [-6.5, 12.9]) and are slightly less likely to engage with feeds (-0.01 engagements decrease, 95\% CIs [-0.06, 0.04])---though the differences are not statistically significant.

While the prompt increases the extent to which participants indicate that they learned something new by browsing the feed (Survey: Q2, 3.2 points increase, 95\% CIs [-8.2, 14.4]), it causes recipients to be slightly less-likely to report that what they saw was different than what they are used to seeing (Survey: Q1, -4.2 points change, 95\% CIs [-15.7, 7.3]), even after controlling for the feed's political alignment (i.e., same vs. opposite). Furthermore, participants who receive the prompt are slightly more likely to report that they understand why others might identify with the opinions shown in the feed (Survey: Q3, 6.7 points increase, 95\% CIs [-5.1, 18.4]). None of these results are statistically significant.  

Interestingly, accounts receiving the prompt treatment appear to be more interested in having a conversation with the feed owner in the future (Survey: Q4, 15.9 points increase, 95\% CIs [2.9, 28.8]). The presence of an effect here is surprising  for two reasons: ($i$) the treatment constituted a minor change in the default experience, and ($ii$) the users were shown the prompt only once: immediately before they load another feed for the first time.  The results suggest the potential power of framing effects when conducting interventions on social media platforms, and how certain types of framing can alter how we perceive others, particularly in the context of an environment like social media that offers limited context about other users.  

%
%

\paragraph{Interaction Effects.} We do not find any significant interaction effects between the two treatments, i.e., the directional effects of administering one treatment (e.g., showing an opposing feed) is consistent across all outcomes, regardless of administering the other (e.g., showing the empathic prompt). Figure~\ref{fig:interaction-effects} shows these results. 

However, by observing Figure~\ref{fig:interaction-effects}, we can see several notable differences in responses between participants who are browsing politically-opposing feeds, depending on whether they first saw the empathic or control prompt.  We use non-linear hypothesis tests~\cite{burkner2017brms} to assess the significance of these differences.  The results reveal that participants who view the empathic prompt followed by a politically-opposed feed are significantly less likely to perceive the feed as different from what they're used to seeing, compared to browsing an opposing feed after seeing the control prompt (10.3 points decrease, CIs [-18.0,-2.5], Figure~\ref{fig:interaction-effects}C).  This finding may help explain part of the results depicted in Figure~\ref{fig:interaction-effects}E: namely, that participants who see the empathic prompt and subsequently browse a politically-opposed feed are significantly more likely to understand why some people might identify with the views shown in the feed, compared to browsing an opposing feed after seeing the control prompt (7.9 points increase, CIs [0.4, 15.3]).  Together, these results suggest that the prompt may be effective in bridging perceived psychological divides and promoting an understanding of why a member of the political outgroup holds the views they do.  

Interestingly, participants who see the empathic prompt and subsequently browse a politically \textit{similar} feed are also more likely to want to have a conversation with the feed owner (15.9 points increase, CIs [2.9, 28.8], Figure~\ref{fig:interaction-effects}F); we do not see this effect for those who view the empathic prompt and then browse opposing feeds, perhaps because there is already a high baseline level of interest in having a conversation with those across the political aisle, and so, the prompt has a limited additional effect.  This high baseline interest is sensible given the purpose of the browser extension and thus likely associated selection biases among participants.

%
%
\section{Discussion}
Our experiment demonstrates that, while exposure to politically-opposed Twitter feeds fuels engagement, it does not enhance participants' understanding of why others might hold opposing views, unless this exposure is framed (a priori) in terms of a familiar experience like disagreeing with a friend.

Information novelty might help explain why politically-opposite feeds increase engagement.  The fact that respondents indicate politically-opposite feeds as different from what they are used to seeing in their own feeds implies these opposing feeds contained novel content.  Prior studies have demonstrated increased user engagement with content that is novel, for example, in the context of online news consumption and sharing~\cite{bergerMilkmanViral}.  Furthermore, as novel information is often surprising, it also attracts more attention~\cite{noveltyAttention,bayesianAttention}.  In the context of social media platforms, this enhanced engagement is encouraging as it demonstrates a consumer appetite for accessing alternative viewpoints and may actually dovetail with key business objectives such as increasing user engagement or time spent on the platform.

Prior work, however, has illuminated that a greater \textit{quantity} of engagement does not always translate into a higher \textit{quality} of  engagement, especially, in the context of politics~\cite{falseNewsPaper}.  In particular, explicit partisan framing has been shown to adversely affect open-mindedness and the civility of ensuing public discourse~\cite{centolaSocialLearning}.  These studies may help illuminate why, on its own, viewing politically-opposite feeds does not promote understanding and bridge empathy gaps between disparate group despite enhancing engagement.

The strong influence of the empathic prompt on promoting understanding can, in part, be explained by framing effects and their ability to alter the nature and quality of engagement and decision-making~\cite{prospectTheory}.  Framing effects are a foundational concept in behavioral science and have been successfully used to drive behavior change across a range of domains like marketing and advertising~\cite{marketingFraming}, energy consumption~\cite{allcottEnergyConservation}, and education~\cite{globalAchievementGap}, to name a few.  Prior work in social psychology and perspective taking sheds some light on the possible drivers of these results.  For example, a study by Lamm et al.~\shortcite{lammEmpathy} involved showing participants videos of facial expressions produced by patients receiving medical treatment, along with whether or not the treatment was effective.  The researchers found that participants who viewed videos of patients undergoing ineffective treatments experienced more distress and less empathy.  In this vein, thinking of the empathic prompt as describing an unsettling encounter that is, ultimately, non-adversarial (the nature of many disagreements with friends) may help explain why prompt recipients were more likely to respond with empathetic attitudes towards politically disparate feed owners.  It is also possible that the friendship framing enhanced perceived commonality between the participant and feed owner.  Indeed, prior work suggests that greater perceived commonality between members of opposing groups may serve as one channel through which to reduce affective polarization~\cite{levenduskyAmericanIdentity, wojcieszak2020affective}, suggesting that the increases in cognitive empathy we observed may offer a step towards reducing this particularly-divisive form of political discordance.

Both the software and framing mechanism we developed in our study are simple to design and deploy, suggesting social media platforms could realistically build them into their platforms if they wish to reduce affective polarization among their users.  For example, platforms might create tools like our extension and make them easily discoverable and accessible to users, who can exercise their individual agency in deciding whether to use them in order to explore viewpoints that differ from their own.  These tools may even be issue-specific: for example, tools that enable users to easily browse the different sides of a trending issue or conversation unfolding on the platform.  They could also enable further filtering by enabling users to browse the viewpoints of specific groups of people vis-a-vis certain issues (e.g., those in rural areas; those in religious groups that differ from their own; etc).  Critically, these tools must be designed---through the use of empathic prompts or other features---to scaffold exposure to diverse views in ways that promote understanding without triggering backfire effects.  Striking this balance will likely require a commitment to continuous iteration and research on the part of platforms.  Our results suggest that making investments in developing these types of features may actually align with business objectives like increasing platform engagement, especially among the subsets of users who take interest in them.

We note that the development of tools and prompts for scaffolding exposure to diverse viewpoints is orthogonal to changes in the underlying algorithms used to personalize content exposure.  We are not arguing against changes to content personalization algorithms---especially given the influence they have on which content people access and the opinions they form---but emphasize that they are  distinct from optional user experiences to illustrate the multiple levers that exist when determining how to promote exposure and reduce negative inter-group affect.  Tight-knit collaborations between platforms and academic researchers from different disciplinary backgrounds may help inform the design and evaluation of experiments that increase our collective understanding of the relative merits and drawbacks of targeting these two (and other) levers for reducing affective polarization online.

Before concluding, we note several limitations in our approach and analysis.  For one, selection bias in participation prevents our ability to generalize findings to wider audiences (e.g., all Twitter users in the U.S., or more broadly, all social media users.\footnote{Though arguably, given the gradual and optional nature in which features like those offered through the extension might be deployed to wider audiences in the future, this selection bias may actually strengthen the external validity of our results.})  Many of these selection biases stem from media coverage of the extension during the specific moment in time that we launched the experiment: in the months following the 2016 US Presidential Election, when there was particularly high interest (especially among left-leaning groups) to better-understand the political philosophies and motivations of those who voted for Donald Trump.  Furthermore, we are also only able to recover political ideologies for a subset of participants, which may further bias towards those who are more likely to share content and be active on Twitter.  Our selection of feeds may have also influenced the results: by selecting users who appear to follow unambiguously left- or right-leaning Twitter accounts, we may have exacerbated the likelihood of inciting backfire effects (as suggested in Garimella et al.~\shortcite{garimellaControversy} and Gillani et al.~\shortcite{socialMirror}) that impede understanding and empathy.  Finally, our research design prevents us from understanding the precise causal mechanisms at play in making the empathic prompt effective in enabling participants to understand why feed owners hold the views they do, although some of the studies highlighted earlier offer hypotheses (e.g., increasing perceived commonality) that may serve as the basis of future experiments.  We believe these limitations offer promising avenues for future work on this emerging topic.

%
%
\section{Conclusion}
In this paper, we describe results from a randomized field experiment conducted on Twitter to investigate how social media platforms might help reduce affective polarization and enhance understanding across political divides.  We deploy the experiment through a Google Chrome extension that enables users to replace their Twitter feeds with a feed from a different politically-active Twitter user.   We find that feeds containing information that contrasts with participants' political views enhance engagement but not an understanding of why someone might hold these divergent views.  On the other hand, prompting users to think of politically-opposing content in the context of a prior disagreement with a friend does increase understanding, but not engagement.  Together, these findings illustrate how social media platforms can take simple steps to reduce affective polarization on their platforms, and how these steps may, in some cases, align with business objectives like enhancing user engagement and time spent on the platform.

There are several opportunities for future work.  Experimenting with different prompts, for example, may shed additional light on which types of re-framing are more or less effective in reducing affective polarization.  These experiments may also help shed light on how social media platforms can adopt and deploy such changes through their platforms.  Furthermore, the fact that many participants indicated a desire to have a conversation with individuals who hold opposing views suggests a future where social media platforms might help extend online conversations into deeper, offline engagements between individuals and/or groups.  While much of the current public discourse attributes worsening affective polarization to the form and function of existing social media platforms, we hope our findings will motivate new efforts to explore how these powerful technologies can help mitigate polarization and promote empathy across divides.

\section{Replication and Supplementary Materials}
\label{sec:replication-and-supmat}
Supplementary materials and the code needed to replicate the analyses presented in this paper can be accessed in the following Github repository: \url{https://github.com/msaveski/twitter_perspective_taking}.

The supplementary materials\footnote{\url{https://github.com/msaveski/twitter_perspective_taking/raw/main/supplementary-material.pdf}} contain: (1) further description and justification of the mixed-effects models used in the main analyses, (2) checks for the robustness of results across different model specifications, (3) checks for covariate balance to ensure that the users assigned to treatment and control have similar background characteristics, and (4) additional analyses of results accounting for selective attrition in the survey responses.

%
%
\fontsize{9.0pt}{10.0pt}  
\selectfont
\bibliography{refs}

\end{document}